\documentclass{article}

\usepackage{PRIMEarxiv}

\usepackage[utf8]{inputenc} 
\usepackage[T1]{fontenc}    
\usepackage{hyperref}       
\usepackage{url}            
\usepackage{booktabs}       
\usepackage{amsfonts}       
\usepackage{nicefrac}       
\usepackage{microtype}      
\usepackage{lipsum}
\usepackage{fancyhdr}       
\usepackage{graphicx}       
\graphicspath{{media/}}     
\usepackage{amsmath}

\usepackage[numbers,sort&compress]{natbib}
\usepackage{cleveref}
\title{Selective Hydrogenation Promotes Anisotropic Thermoelectric Properties of TPDH-Graphene
}

\author{
  Caique Campos de Oliveira\\
  Center for Natural and Human Sciences (CCNH) \\
  Federal University of ABC (UFABC) \\
  Santo André - SP, 09210-170, Brazil.\\
  \texttt{caique.campos@aluno.ufabc.edu.br} \\
   \And
  Douglas Soares Galvao \\
  Physics Institute Gleb Wataghin (IFGW) \\
  State University of Campinas (UNICAMP) \\
  Campinas/SP, Brazil\\
  \texttt{galvao@ifi.unicamp.br} \\
  \And
  Pedro Alves da Silva Autreto\\
  Center for Natural and Human Sciences (CCNH) \\
  Federal University of ABC (UFABC) \\
  Santo André - SP, 09210-170, Brazil.\\
  \texttt{pedro.autreto@ufabc.edu.br} \\
}

\begin{document}

\maketitle

\begin{abstract}
We have combined DFT calculations with the Boltzmann semiclassical transport theory to investigate the effect of selective hydrogenation on the thermoelectric properties of tetra-penta-deca-hexagonal graphene (TPDH-gr), a recently proposed new 2D carbon allotrope. Our results show that the Seebeck coefficient is enhanced after hydrogenation. The conductivity along the x direction is increased almost eight times while being almost suppressed along the y direction. This behavior can be understood in terms of the electronic structure changes due to the appearance of a Dirac-like cone after the selective hydrogenation. Consistent with the literature, the electronic contribution to thermal conductivity displays the same qualitative behavior as the conductivity, as expected from the Wiedemann-Franz law. The increase in thermal conductivity with temperature limits the material's power factor. The significant increase in the Seebeck coefficient and conductivity increases also contribute to the thermal conductivity increase. These results show that hydrogenation is an effective method to improve the TPDH-gr thermoelectric properties, and this carbon allotrope can be an effective material for thermoelectric applications.

\end{abstract}

\section{Introduction}
Anisotropic physical properties generally originate from structural low symmetry \cite{Zhao2020}. Orthorhombic, monoclinic, and triclinic lattices are likely to exhibit directional-dependent properties due to the non-symmetrical chemical environment of the atoms in these structures \cite{Zhao2021}. Examples of such materials include low-symmetry chalcogenides like Re$X_{2}$ (X = S or Se) with anisotropic optical properties \cite{Echeverry2018}, materials with non-trivial electronic topology (e.g., Weyl-semimetals, such as Nb(S, As)) \cite{Zhou2019}, and bismuth tellurides \cite{Witting2019}. Metal-free materials can also exhibit anisotropy. In black phosphorus (BP), the different hybridization of each P atom results in anisotropic electrical properties \cite{Xia2014}. Borophenes, 2D boron monolayers \cite{Piazza2014, Feng2016} exhibit anisotropic mechanical \cite{Zhang2017} and thermal \cite{Zhou2017} properties. 

Directional-dependent properties are highly desirable in some energy harvesting applications, including thermal energy recycling through Seebeck or Peltier effects \cite{Burkov1995, Chakraborty2018, Yun2021}. The rise of two-dimensional (2D) compounds opened new perspectives in the development of next-generation nanodevices. In contrast to 3D or “bulk” materials, their 2D counterparts exhibit reduced thickness with a large surface-to-volume ratio, giving rise to electronic confinement effects that can be engineered to tune structural, electronic, optical, and thermoelectric properties \cite{Zhao2021, Dresselhaus2007}. 2D carbon allotropes represent an exciting group of these materials due to their unique structural and electronic properties. The hybridization of valence orbitals in carbon enables the existence of topologically different allotropes, such as graphene \cite{Novoselov2004}, graphynes \cite{Ivanovskii2013, Desyatkin2022}, biphenylenes \cite{Fan2021}, among others.  

From an environmental perspective, carbon-based materials are attractive due to their high availability, non-toxicity, and eco-friendliness \cite{Yun2021}. Their physical properties are also technologically attractive. Graphene, for example, exhibits excellent mechanical, optical, electrical, and thermal properties suitable for electronics, optoelectronics, and sensing \cite{Sang2019}. However, its gapless electronic behavior and exceptionally high thermal conductance hinder its thermoelectric applications \cite{Tabitha2018, Dollfus2015}. The enhancement of thermoelectric properties of materials requires the modulation of electrical and thermal conduction \cite{Wolf2019} and can be achieved by nanostructuring, doping/defects \cite{Tabitha2018} and functionalization \cite{Wu2016, Li2019}. 

\begin{figure}[t!]
 \centering
  \includegraphics[scale=0.93]{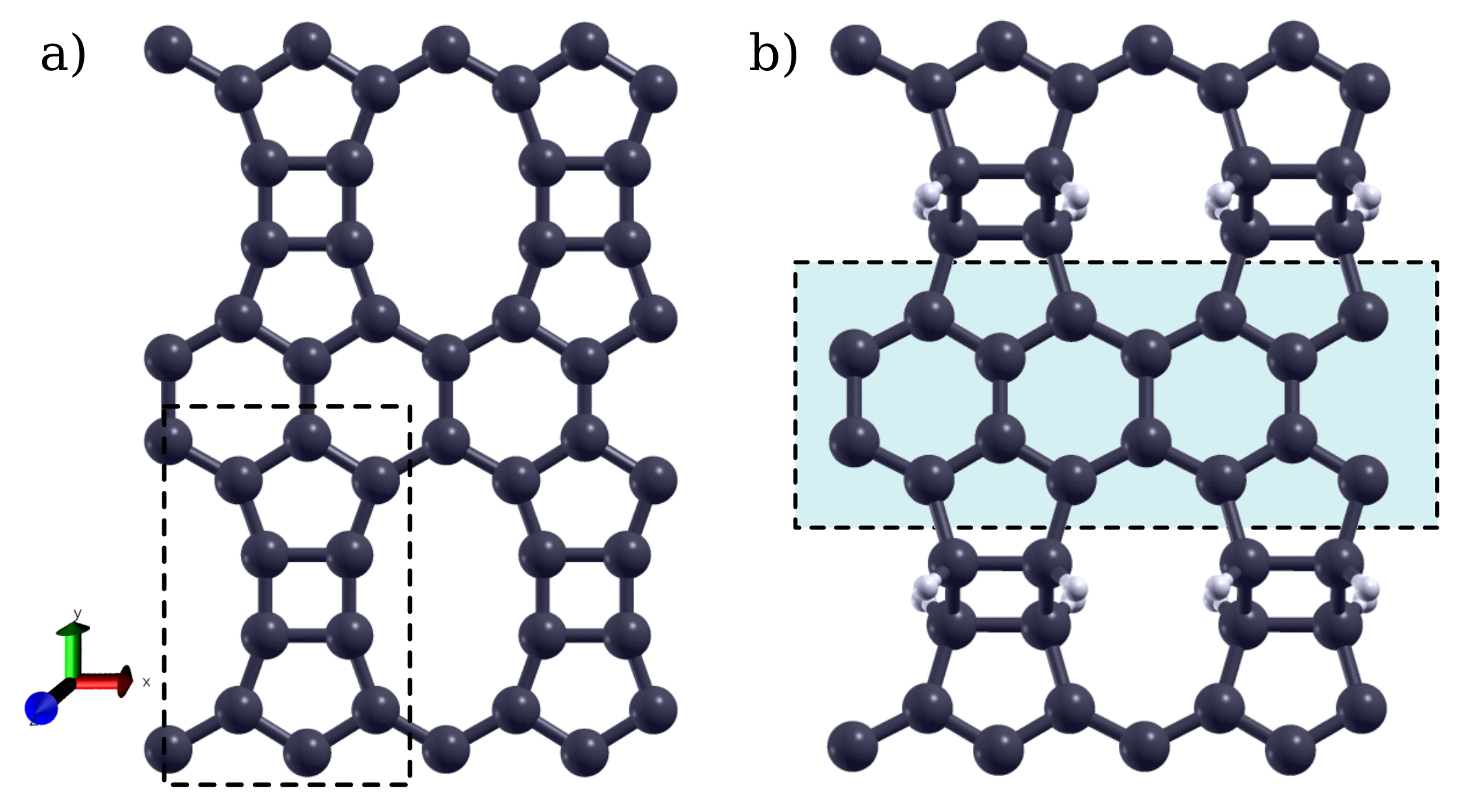}
  \caption{a) A 2x2 TPDH-gr supercell. b) a 2x2 hTPDH-gr. The hexagonal carbon rings along the x direction are highlighted in the dashed light green rectangle.}
  \label{fgr:structure}
\end{figure}

Recently, novel carbon 2D allotropes have been experimentally realized, such as graphynes \cite{Ivanovskii2013, Desyatkin2022} and biphenylenes \cite{Fan2021}. Biphenylenes are of particular interest because they combine different carbon rings that can lead to low-symmetry anisotropic systems. Tetra-penta-deca-hexagonal graphene (TPDH-gr, Figure 1(a)) is another example of such structures. TPDH-gr exhibits in-plane elastic anisotropy and significant optical adsorption in short wavelength range, making it promising as a coating material for UV protection \cite{BHATTACHARYA2021}. However, TPDH-gr presents semimetallic behavior, which is known to limit thermoelectric performance. Fortunately, this limitation can be overcome by attaching hydrogen adatoms to the structures to tune the electronic properties. This methodology has been successfully applied to other 2D metal-free structures, including carbon, silicon, and germanium-like ones \cite{Sahin2015, Trivedi:2014:1546-1955:781}.  

We have recently shown that the hydrogenation of TPDH-gr occurs mainly at the tetragonal rings, producing transversal lines of hydrogen atoms with non-hydrogenated hexagon rings along the x-direction, as highlighted in Figure 1(b). These structural changes result in electronic changes, giving rise to Dirac’s cone features \cite{Oliveira2023}. However, this study was limited to investigating the structural and electronic changes. In this work, we applied first-principles calculations (Density Functional Theory (DFT)) together with the Boltzmann semiclassical transport equation to investigate the effects of hydrogenation on the transport properties of pristine and hydrogenated TPDH-gr. We also applied the Boltzmann transport theory to calculate the thermoelectric properties of pristine and hydrogenated TPDH-gr. Our results show that both structures present significant anisotropic behavior, and selective hydrogenation enhances this behavior by almost suppressing the electrical conduction along particular directions.

\section{Computational Methods}
First-principles calculations were carried out using density functional theory as implemented in the Quantum ESPRESSO package \cite{Giannozzi2017}. Core electrons were approximated by the pseudopotentials (pp) method, using the solid standard state (SSSP) library \cite{Prandini2018}. Kohn-sham orbitals were expanded using a plane wave basis set imposing a threshold of $60$ Ry for the kinetic energy and a $480$ Ry one for the electronic density, as suggested by the pp efficiency set of SSSP. The Brillouin zone was sampled by a $(16 \times 16 \times 1)$ $6 \times 6 \times 1$ k-point mesh for (non)self-consistent calculations following the Monkhorst-Pack scheme \cite{Monkhorst1976}. The threshold for self-consistency of electronic density was set to $10^{-7}$ Ry, while the energy and forces between two consecutive ionic steps were minimized until $10^{-4}$ Ry and $10^{-3}$ Ry/Bohr, respectively. The exchange and correlation interactions were approximated by the generalized gradient parametrized by Perdew, Burke, and Ernzerhoff  (GGA-PBE) \cite{PBE1996}.  

The electronic transport coefficients were calculated using Boltzmann semiclassical transport equations within the relaxation time approximation. The conductivity ($\sigma$), Seebeck coefficient ($S$), and electrical contribution to thermal conductivity ($\kappa_{e}$) tensors were obtained from analytical expressions as functions of temperature ($T$) and chemical potential ($\mu$) using the transport kernel ($\Xi_{ij}(\epsilon)$): 

\begin{align}
    \sigma_{ij}(T, \mu) = \frac{1}{\Omega} \int 
 \Xi_{ij}(\epsilon) \left[ \frac{-\partial f_{\mu}(T, \epsilon)}{\partial \epsilon}\right]d\epsilon \\
  (\kappa_{e})_{ij}(T, \mu) = \frac{1}{e^{2}T\Omega} \int 
 \Xi_{ij}(\epsilon - \mu)^{2} \left[ \frac{-\partial f_{\mu}(T, \epsilon)}{\partial \epsilon}\right]d\epsilon \\
  S_{ij} = \frac{1}{eT\Omega \sigma_{ij}} \int 
 \Xi_{ij}(\epsilon - \mu) \left[ \frac{-\partial f_{\mu}(T, \epsilon)}{\partial \epsilon}\right]d\epsilon
\end{align}
Where e is the electrical charge, $\Omega$ is the cell total volume, $\epsilon$ is the electronic band energy, and $f_{\mu}(T, \epsilon)$ is the carrier distribution function. The $i$ and $j$ indexes represent the cartesian components. BoltzTraP code \cite{MADSEN2006} performs a Fourier expansion of the band energies, interpolating these values to obtain an analytical function that is used to evaluate the thermoelectric coefficients $S$, $\sigma/\tau$, and $\kappa_{e}/\tau$, where $\tau$ is the time relaxation parameter. These coefficients are calculated within the rigid band approximation, assuming that the band structure does not significantly change for different doping levels \cite{MADSEN2006, Li2019, TROMER2021}.

\section{Results and Discussion}
\subsection{Seebeck Coefficient}

\begin{figure}[t!]
  \centering
  \includegraphics[scale=0.53]{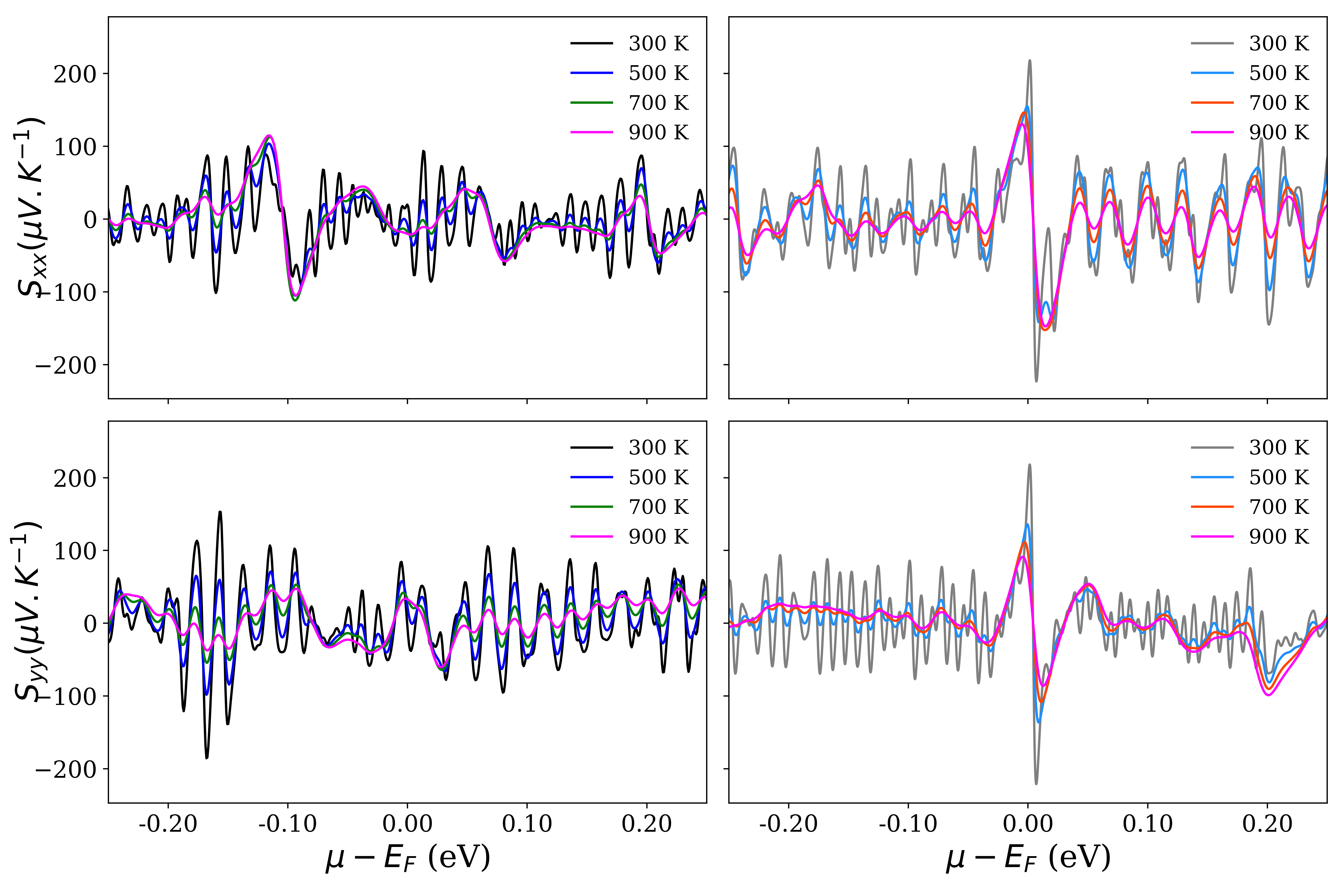}
  \caption{a) Seebeck coefficient as a function of the chemical potential ($\mu$) shifted relative to the Fermi energy tensor components for the x (top) and y (bottom) directions, for pristine (left) and hydrogenated (right) TPDH-gr at different temperature.}
  \label{fgr:seebeck_coeff}
\end{figure}

The conversion of waste heat into electricity can be achieved by exploiting the Seebeck or Peltier effects. The Seebeck coefficient ($S$) represents the maximum potential difference created in the material when a temperature gradient is present \cite{Yun2021}. The x and y components of the S tensor for pristine and hydrogenated TPDH-gr are presented in Figure 2. Pristine TPDH-gr displays a small $S$ coefficient around the Fermi level for both directions. This is typical for metals due to the symmetry of electrons and holes caused by the absence of an electronic band gap \cite{Markov2019}. Also, as the temperature increases, the profile of the curves becomes very similar, indicating that the Seebeck coefficient reaches a peak value. Analyzing the $S$ tensor as a function of the temperature ($T$) (Fig S.1), it is possible to infer that the $x$ component of $S$ for pristine TPDH-gr is zero for $T = 700$ K, while the y-component reaches a limit value of $–80$ $\mu$V/K at $T = 500$ K. The corresponding curves for hydrogenated TPDH-gr are also different for the x and y directions, as expected from the material structural anisotropy. When the structure is hydrogenated, the Seebeck coefficient exhibits a peak near the Fermi level, reaching $200$ $\mu$V/K at $300$ K for both directions. The inverse sign of the peaks indicates that the charge carriers are holes (for the positive peak) and electrons (for the negative one) \cite{TROMER2021}. For both analyzed directions, the peaks’ amplitude decreases with increasing temperature, although this behavior is more pronounced for the y direction, showing that the anisotropic behavior is preserved. The decrease can be attributed to the increase in thermal energy, similar to the observed in other materials \cite{Pakizeh2019}. 

\subsection{Electric and Electronic contribution to Thermal Conduction}

In Figure 3, we present the results for the electrical and thermal conductivities as a function of temperature for the pristine and hydrogenated TPDH-gr. From Figure 3, the anisotropic behavior is evident. The conductivity along the x direction is constant (roughly $1 \times 10^{19}$ ($\Omega.m.s)^{-1}$). In contrast, the y component increases with temperature: for $T = 300$K, $\sigma_{y} \approx 3 \times 10^{19}$ $(\Omega.m.s)^{-1}$ reaching $\sigma_{y} = 4.5 \times 10^{19}$ $(\Omega.m.s)^{-1}$ for $T > 700$ K. A similar behavior is observed for the thermal conductivity.  

These results can be explained in terms of the electronic structure (presented in Fig. S.2). Pristine TPDH-gr exhibits an electronic band structure with the valence band crossing the Fermi energy level along the $\Gamma \xrightarrow{} Y$ (Figure 4(b)). From the Brillouin zone of the orthorhombic cell (see Figure 4), it becomes clear that this path is directly related to the y direction. After hydrogenation, this behavior changes completely. The conductivity is almost eight times higher along the x direction compared to the pristine structure, ranging from 
$\sigma_{x} = 7.0 \times 10^{19}$
($T = 300$ K) to $\sigma_{x} = 8.0 \times 10^{19}$ $(\Omega.m.s)^{-1}$ at $T = 900$ K. $\sigma_{y}$ is substantially reduced, exhibiting an almost constant value of 
$0.5 \times 10^{19} \quad (\Omega.m.s)^{-1}$. This results in a variation in the ratio of conductivities ($\sigma_x/\sigma_y$) ranging between 14 and 16 as the temperature rises.

We can understand these changes by analyzing the electronic structure of the hydrogenated system (Figure 4 and Fig. S2.b). We can see that there is a Dirac-type cone along the $\Gamma \xrightarrow{} X$ path, while at the $\Gamma \xrightarrow{} Y$ one, the structure displays only a small bandgap at the $Y$ symmetry point. Also, these changes are consistent with the electronic structures of nanoribbons studied by Bathacharya and Jana \cite{BHATTACHARYA2021}. The nanoribbons along the y direction (involving tetra and pentagonal rings) are semiconductors, while the nanoribbons along the x direction (involving pentagonal rings, see Figure 1(b)) are metallic. Topologically, we can consider that after hydrogenation, there is a confinement of the electrons through the homogeneous stripes composed of hexagonal rings (Figure 1(b)), which are expected to exhibit a higher conductivity when compared to the y direction (inhomogeneous ring arrays). 

\begin{figure}[t!]
  \centering
  \includegraphics[scale=0.53]{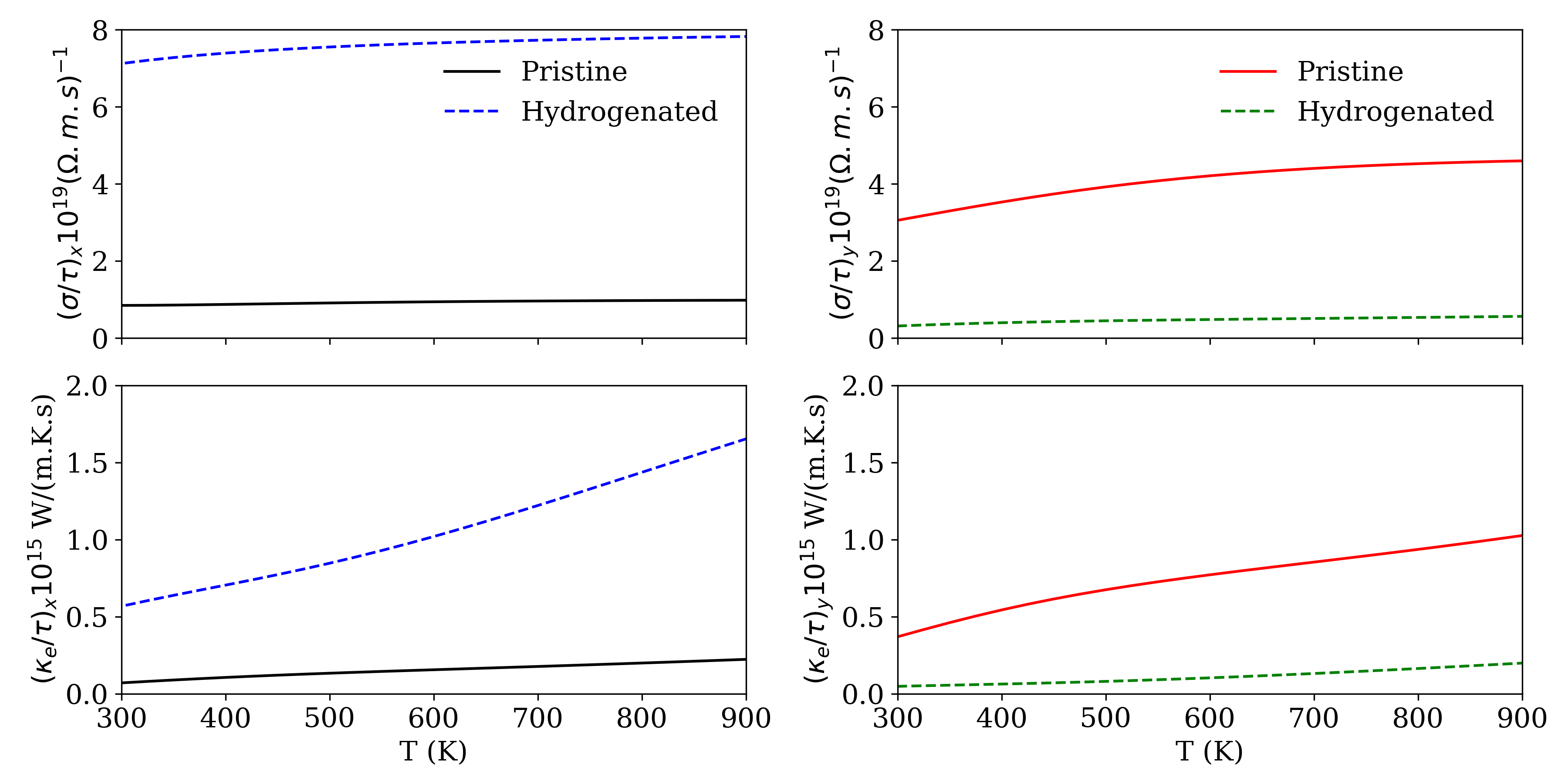}
  \caption{a) Electrical (top) and thermal (bottom) conductivity tensor components along the x (left) and y (right) directions for pristine (solid) and hydrogenated (dashed) TPDH-gr.}
  \label{fgr:conductivities}
\end{figure}

The results for the electronic contribution to the thermal conductivity ($\kappa_{e}$)
show a temperature dependence for both components. An increase in thermal conductivity is expected due to increased thermal energy. The x direction also shows a lower conductivity when compared to the y direction. When the structure is hydrogenated, $\kappa_{ex}$ increases while $\kappa_{ey}$drops substantially. This indicates that hydrogenation also promotes thermal transport along the x direction while suppressing the transversal heat conduction. These results are the opposite of the hydrogenation of graphene but agree with the results for penta-graphene, in which the enhancement in the thermal conductivity was attributed to weaker phonon scattering due to larger bond anharmonicity \cite{Wu2016}. Both components display similar behavior relative to the electric conductivity ($\sigma$), an expected result since these quantities are related by the Wiedemann-Franz law \cite{ALAVIRAD2020}. The anisotropy of hydrogenated TPDH-gr could be useful in electronic applications in which directional electric currents are applied. 

\begin{figure}[t!]
  \centering
  \includegraphics[scale=0.75]{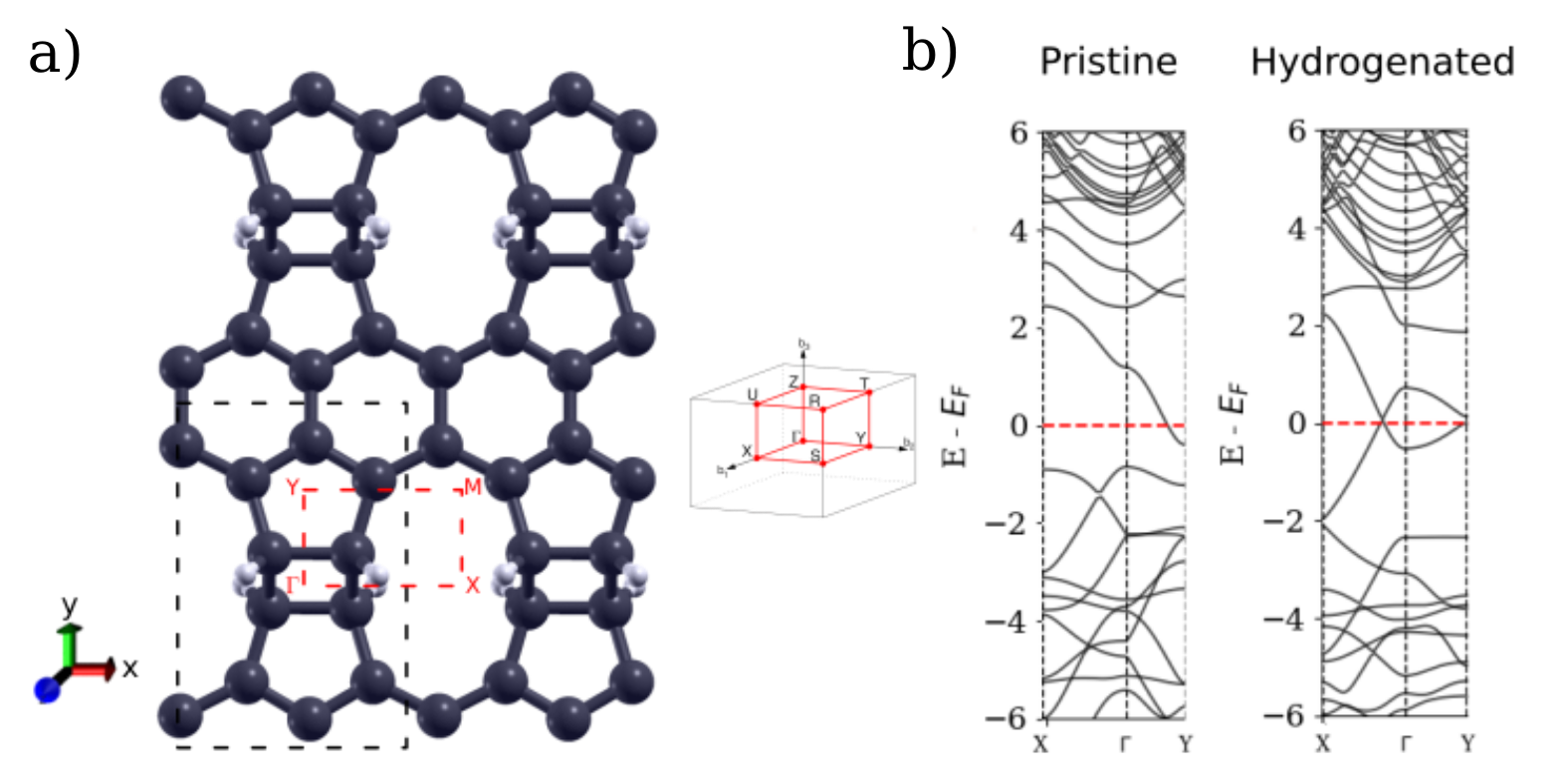}
  \caption{a) A hTPDH-gr 2x2 supercell. The black-dashed lines indicate the corresponding unit cell, while the red-dashed ones indicate the high-symmetry points ($\Gamma$, $X$, $Y$, and $M$) in the first Brillouin zone of the orthorhombic lattice system (inset). (b) Electronic band structure for pristine and hydrogenated TPDH-gr along the $X \xrightarrow{} \Gamma \xrightarrow{} Y$ path high-symmetry points.}
  \label{fgr:BZ-inset}
\end{figure}

\subsection{Power Factor}

The thermoelectric conversion capacity, representing the maximum power available, can be expressed by the power factor (PF) given by the product $S^{2}\sigma$. The x and y components of the PF are presented in Figure 5.  

As expected, pristine TPDH-gr presents a poor thermoelectric conversion capacity. It has an almost zero PF along the x direction and a maximum of $5x \times 10^{10} \quad W/(mk^{2}s)$ along the y direction. For comparison, the PF of $6\%$ compressed n-type doped hydrogenated germanene can reach $57.21 \times 10^{10} \quad W/(mK^{2}s)$ at $T = 800$ K \cite{ALAVIRAD2020}. The PF of Zintl phase (Sr,Ba)$Ag_{2}SeTe$ compounds can reach 
$17.1 \times 10^{10} \quad W/(mK^{2}s)$ at $300 K$ reaching almost $70 \times 10^{10} \quad W/(mK²s)$ at $T = 1200$ K \cite{BEHERA2022}. After hydrogenation, the corresponding PF x-component is significantly enhanced, reaching $30 \times 10^{10} \quad W/(mK²s)$ at $T \approx 530K$, while the y component is almost suppressed (near zero PF value).

\begin{figure}[t!]
  \centering
  \includegraphics[scale=0.6]{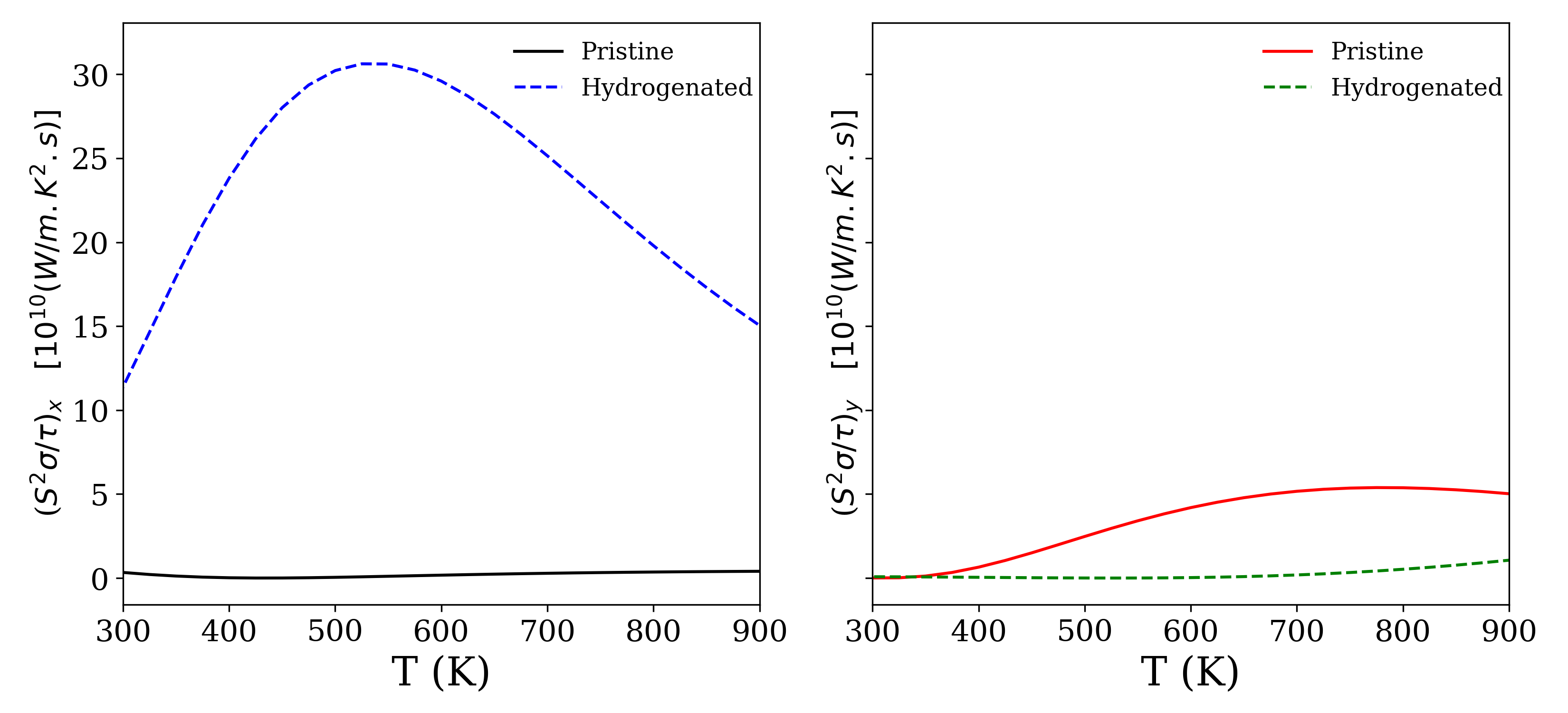}
  \caption{a) Power factor components for x and y directions. Solid and dashed curves refer to pristine and hydrogenated results, respectively.}
  \label{fgr:PF}
\end{figure}

These results can be understood in terms of the changes in the electrical conductivity caused by hydrogenation. There are two main contributing factors to the thermoelectric properties of a material: the Seebeck coefficient and the lattice thermal conduction ($\kappa = \kappa_{ph} + \kappa_{e}$) \cite{Markov2019}. While hydrogenation was shown to promote the thermoelectric conversion by increasing $S$, $\kappa_{e}$ was also increased, as seen by the increase in the x component (Figure 3(c, d)). Therefore, to further improve the TPDH-gr thermoelectric properties, there are two directions to follow: (i) further enhancement of the Seebeck coefficient, which could be achieved by increasing the hydrogen-carbon ratio, thus possibly increasing the electronic bandgap. (ii) decrease the thermal lattice conductivity by increasing phonon scattering. Combining these two strategies, enhancing the TPDH-gr figure of merit (ZT) should be possible.

\section{Conclusions}
In this work, we have combined DFT calculations with the Boltzmann semiclassical transport theory to investigate the effect of hydrogenation on the thermoelectric properties of TPDH-gr. Our results show that TPDH-gr thermoelectric properties are intrinsically anisotropic due to the low symmetry of the lattice system. Furthermore, the Seebeck coefficient is enhanced after hydrogenation due to the changes in the material's electronic structure after selective hydrogenation of the tetragonal rings. Interestingly, the conductivity along the x direction is increased almost eight times while being almost suppressed along the y direction. This behavior can be understood in terms of the electronic structure changes due to the appearance of a Dirac-like cone after the selective hydrogenation. Consistent with the literature, the electronic contribution to thermal conductivity displays the same qualitative behavior as the conductivity, as expected from the Wiedemann-Franz law. The increase of thermal conductivity with temperature limits the material's power factor. The significant increase in the Seebeck coefficient and conductivity increases also contribute to the thermal conductivity increase. These results show that hydrogenation is an effective method to improve the TPDH-gr thermoelectric properties, and this carbon allotrope can be an effective material for thermoelectric applications. Considering the recent advances in the synthesis of new 2D carbon allotropes \cite{Fan2021, Desyatkin2022, Toh2020}, including structures with some of the same TPDH-gr carbon rings \cite{Fan2021}, the TPDH-gr synthesis can be considered feasible with our present-day synthesis capabilities.


\section*{Conflicts of interest}

There are no conflicts to declare.

\section*{Acknowledgements}


CCO thanks PRH.49 (PRH-ANP UFABC) for funding and UFABC Multiuser Computational Center (CCM-UFABC) for computational resources provided. DSG thanks the Center for Computational Engineering \& Sciences (CCES) at Unicamp for financial support through the FAPESP/CEPID Grant 2013/08293-7 and PASA to CNPq (Grant 308428/2022-6).





\bibliography{main} 

\end{document}